# Perspective on Atomic-Resolution Vibrational Electron Energy-Loss Spectroscopy

Benedikt Haas[1,*], Christoph T. Koch[1], Peter Rez[2]

[1] Department of Physics & Center for the Science of Materials Berlin, Humboldt-Universität zu Berlin, 12489 Berlin, Germany
[2] Department of Physics, Arizona State University, Tempe, AZ 85287-1504, USA
* Corresponding author: haas@physik.hu-berlin.de

ABSTRACT

Instrumentation developments in electron energy-loss spectroscopy (EELS) in the scanning transmission electron microscope (STEM) one decade ago paved the way for combining milli-electronvolt energy resolution in spectroscopy with Ångström-sized electron probes, unlocking unexplored realms for solid state physics at the nanometer-scale. The fundamental understanding of the scattering processes involved has made it possible to separate atomically localized signals, providing insights into vibrations at the atomic scale. In this Letter, we outline the experimental, conceptual and theoretical advances in this field and also make comparisons with tip-based optical techniques before discussing future perspectives of this exciting technique. Optimization of dark-field signal collection will play a fundamental role in making this technique more widely applicable to a range of materials problems. The benefits of adding momentum-resolution will also be discussed and a powerful acquisition scheme proposed.

MAIN

Vibrational spectroscopy using electrons was successfully implemented for broad (~0.1 mm) beams in scattering experiments in the 1960s [1] using a double Wien-filter for energy selection and later with a more localized beam of < 0.1 μm [2]. Electron spectroscopy in electron microscopes capable of atomic resolution was limited to > 40 meV at best for monochromators available before 2014 [3,4]. The introduction of the ground-potential monochromator by Krivanek et al. made it possible to combine the high resolution imaging and diffraction capabilities of the electron microscope with the meV energy resolution needed to detect vibrational modes [5]. A discussion of the hardware developments enabling this vibrational spectro-microscopy can be found elsewhere [6].

The spectrometer developed by Krivanek et al. can also accept a very large range of scattering angles, making it possible to either efficiently collect the EELS signal for an atomically resolving probe of > 30 mrad convergence angle, or record momentum-resolved vibrational spectra in parallel [7]. The latter technique is especially relevant for studying the dispersion of phonons – the collective vibrations of atoms in crystals. As spatial and momentum resolution are linked by Heisenberg's uncertainty principle, high momentum resolution is not possible in combination with atomic spatial resolution and we will not discuss dispersion-resolved experiments here. However, we will investigate certain anisotropies in atomic resolution vibrational EELS resulting from momentum transfer along different crystallographic directions in the outlook.

**Imaging in The Electron Microscope**

High resolution imaging in electron microscopy is either performed in the conventional transmission electron microscope (CTEM) or the scanning transmission electron microscope (STEM). Image formation in the CTEM follows the same principles as image formation in a (conventional) optical microscope. A condenser system produces parallel illumination of the area of interest that is then imaged by the objective lens with further magnification of the intermediate image via projector lenses to give an image on a screen or a 2D detector. In STEM a small probe, that can be of atomic dimension, is formed and scanned across the specimen and a signal such as the scattering onto an annular dark-field detector, is recorded sequentially. For a more in-depth introduction to basic TEM techniques, the reader is referred to standard textbooks such as by Williams and Carter [8]. So far, high-resolution vibrational spectroscopy has only been demonstrated in STEM-based EELS, as energy-filtered conventional CTEM (EFTEM) with a few meV-sized energy-selecting slit would be highly dose inefficient, because almost all scattered electrons would be blocked by the slit and only a tiny fraction of them would be recorded. Therefore, the article focuses exclusively on STEM-EELS.

In STEM the electrons producing the image can typically be considered to originate from a point source since the probe is formed by focusing electrons emitted from a small field emission tip, which is often further demagnified to a diameter smaller than 1 Å. The electron wave can thus be approximated as an ideal spherical wave. If the electron phase is flat over the angular range of the probe (typically defined as deviations in the path length of less than a quarter of the wavelength), then the resolution is limited only by the aperture size. With correction of $3^{rd}$ and $5^{th}$ order aberrations of the objective or focusing lens it is possible to form probes with 40 mrad convergence semi-angle at 60 kV with flat phase distribution. This corresponds to 13.7 $nm^{-1}$ in reciprocal space or a probe size diameter of 0.073 nm in real space, absent higher order aberrations or other influences such as instabilities in lens currents, high tension, or external fields.

To achieve atomic resolution in vibrational STEM-EELS one starts with a probe where the geometric aberrations have been corrected and the convergence angle is sufficient to produce an atom-sized probe on the sample (cf. Fig. 1 a). This probe must be monochromated, since its initial energy distribution limits the attainable energy resolution. The transmitted and scattered electrons, the trajectory of which deviates from the optic axis of the microscope up to the limits defined by the spectrometer entrance aperture, are energy-dispersed using a magnetic prism and collected in parallel on a pixelated detector to obtain a spectrum for each scan point (cf. Fig. 1 a). Efficient signal collection by the spectrometer is imperative as, due to the necessary demagnification of the source, the beam current in atomically-resolved STEM is low (on the order of 100 pA, even lower without aberration corrector) and most of this current is blocked by the monochromator. To optimize the signal, the collection aperture or spectrometer entrance aperture should be at least as large as the probe forming or objective aperture. These high angles pose constraints on residual spectrometer aberrations, an aspect that will be discussed later. It should be noted that aberrations present at the energy-selecting slit of the monochromator (often called 'slit aberrations') stemming from the optics before the slit limit the achievable energy resolution in the same way that condenser aberrations in the sample plane limit the achievable spot size and thus spatial resolution. Therefore, the slit aberrations must be corrected using suitable optics before the slit.
For a typical phonon mapping experiment, on the order of 98 % of electrons are discarded by the monochromator when producing a source with an energy width of 6 meV, starting from the spectrum of a cold field emission source that is about 300 meV wide. The attainable energy resolution, which is not only defined by the monochromation but also limited by spectrometer aberrations, tends to be smaller for reduced beam energy. This can be rationalized by approximately constant relative noise sources, e.g. ripples in currents j (of optical elements) from power supplies with relatively constant Δj/j, which lead to smaller absolute noise Δj for lower currents j as typically used at lower electron energies. However, at the same time, the achievable spatial resolution improves for higher electron energy, so a compromise must be made. In practice 60 keV has proven to be a suitable electron energy, as energy resolution does not improve very substantially below this value, but spatial resolution deteriorates quickly and extremely thin samples are required when working at accelerating voltages below 60 kV.

**Dipole and Impact Scattering**

To form an image showing individual atoms not only must the probe be of atomic dimensions but the scattering must also be localized to atomic sites. The signal picked up by a high-angle annular dark-field (HAADF) detector is dominated by thermal diffuse scattering (TDS), a process in which the incident electron has either created or annihilated one or even multiple vibrational modes. The intensity on the detector can be calculated from an Einstein model similar to that used for computing the Debye-Waller factor [9]. An alternative is to introduce displacements of the atoms, which could correspond to actual phonon modes, the so-called frozen phonon (FP) approximation [10], that relies on the fact that the time the electron takes to traverse the specimen is small in comparison to a phonon oscillation period. The result is a redistribution of intensity from the diffraction discs produced by purely elastic scattering into various angles, giving rise to TDS. Correlations in the atomic displacements in the FP approximation from using actual phonon modes lead to a fine structure in the TDS that is closer to the experimental observation than when applying the Einstein model, but the difference is small when integrating over the HAADF detector [11].

To explore the different possibilities of scattering from individual phonons it is instructive to consider the case of a single bond between two atoms, denoted in the following by the indices 1 and 2. The general expression for the scattering cross section for the creation of a vibration is

$$\frac{d\sigma}{d\Omega} = \left| \mathbf{q} \cdot \mathbf{e} \sqrt{\frac{\hbar(N_0(\omega(q)) + 1)}{2\omega(q)}} \gamma \left( \sqrt{\frac{m_2}{m_1}} \frac{f_{el;1}(q)}{\sqrt{m_1 + m_2}} \pm \sqrt{\frac{m_1}{m_2}} \frac{f_{el;2}(q)}{\sqrt{m_1 + m_2}} \right) \right|^2$$

where $\omega(q)$ is the frequency, $m_1$ and $m_2$ are the masses of the atoms, $N_o(\omega)$ is the Bose Einstein factor, **q** is the scattering wave vector, **e** the polarization direction, $\gamma$ the relativistic correction factor and $f_{el;n}$ is the electron scattering factor for atom 1 or 2. There are two possibilities; either the atoms vibrate in phase which corresponds to the + sign, or they vibrate in antiphase which corresponds to the – sign. In a solid the in-phase vibrations of two distinct atoms in the unit cell would correspond to an acoustic phonon and the antiphase vibrations to an optic phonon.

Let us now also allow for a charge transfer such that one atom carries a charge of $+\Delta\rho$ and the other atom is negatively charged by $-\Delta\rho$.

$$f_{el}^{ion}(q) = \frac{2\Delta\rho}{a_0 q^2} + f_{el}^{neut}(q)$$

where $a_0$ is the Bohr radius. The dipolar part of the cross section for the optic phonon is now

$$\frac{d\sigma}{d\Omega} = \frac{4\gamma^2}{a_0^2} \frac{\Delta\rho^2 \cos^2\psi}{q^2} \left(\frac{\hbar(N_0(\omega(q)) + 1)}{2\omega(q)}\right) \left(\frac{m_1^2 + m_2^2}{m_1 m_2 (m_1 + m_2)}\right)$$

where $\psi$ is the angle with respect to the bond direction. This is analogous to infrared absorption [12] in that there is an induced oscillating dipole.
The scattering wave vector is made up from two parts one parallel to the incident electron direction $k\theta_E$ and the other perpendicular to that direction

$$q^2 = k^2(\theta^2 + \theta_E^2)$$

$$\theta_E = \frac{\hbar\omega}{m_0 v^2}$$

Here, θ is the scattering angle perpendicular to the incident beam direction, $\theta_E$ is the characteristic scattering angle, $m_0$ the electron rest mass and v the electron velocity. Since optic phonon energies are very small, of the order 100 meV, the characteristic scattering angle for a primary electron energy of 60 keV electrons, which is typically used, is of the order a few microradians. From the uncertainty principle a small spread in momentum space is equivalent to a large spread in real space and "delocalization" over distances of hundreds of nm is possible.

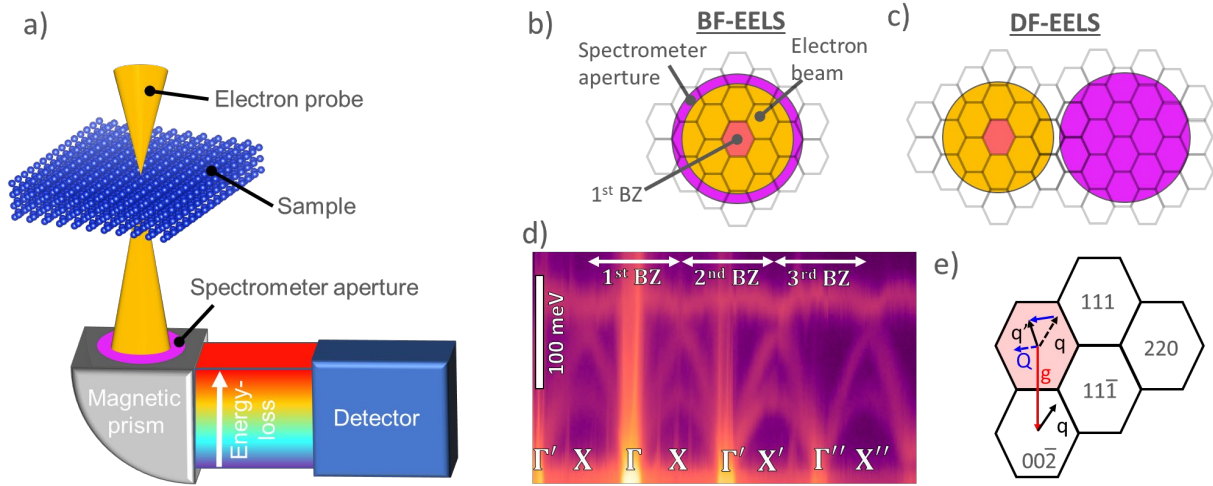

*Figure 1: a) Simplified sketch of the experimental setup for atomic resolution vibrational EELS. Reprinted (adapted) with permission from Haas et al. [13]. Copyright 2023 American Chemical Society. b) Geometry of a typical atomic resolution BF phonon EELS experiment with the grid representing Si Brillouin zones in the [110] zone axis. c) Sketch of the setup for DF-EELS in comparison. d) Experimental phonon dispersion of diamond, showing that the intensity first increases with scattering angle due to dominant Umklapp scattering. e) Sketch of different scattering vectors within the Brillouin zones (contrary to textbook solid state physics but as is common within the electron microscopy, we call the neighboring BZs according to their distance the $n^{th}$ BZs). q is a wavevector in the 1st BZ for the incident electrons. Since the incident beam convergence is large enough to encompass the 00-2 BZ an Umklapp vector g is shown bringing it back to the 1st BZ. q' is wavevector in the 1st BZ for the scattered electrons and Q is the net change in wave vector referred to the 1st BZ which is relevant for the phonon energy.*

The initial nanoscopic vibrational EELS experiments by Krivanek et al. already indicated that the signal was not localized to just the area of the electron probe but delocalized on the order of tens of nanometers [5]. While the possibility to probe material adjacent to the electron probe without exposing it directly to the fast electrons (called "aloof" geometry) was intriguing and later used by Rez et al. to demonstrate the damage-free study of biological materials [14], the consequences for the spatial resolution were rather disappointing. Several studies have been published by now on the theoretical description of aloof EELS and the consequences for damage and how signal strength depends on the impact parameter (distance between beam and probed location) for different geometries, e.g. by Egerton [15].

Although the dipole signal is delocalized over considerable distances, the signal that does not arise from exciting or annihilating a dipole oscillation, called impact scattering, is localized to atomic sites. It can thus, in principle, be used to form images at atomic resolution. This was already recognized in a paper preceding the instrumental development of vibrational electron spectro-microscopy which showed simulations of lattice fringe profiles for phonon scattering perpendicular to a systematic line[10]. At the time of the first experiments, Lugg et al. argued that, since atomic resolution was obviously possible in HAADF imaging which relies on the high degree of localization of high-angle thermal diffuse scattering, it should thus also be possible for vibrational EELS. Scattering from acoustic phonons with no oscillating dipole is clearly an example of impact scattering, but not all signal from optic phonons is dipole scattering. Dwyer et al. showed in 2016 that besides the delocalized dipole scattering there is also a localized phonon signal and demonstrated a spatial resolution of about one nanometer [16]. If dipole scattering is analogous to infrared absorption, impact scattering is comparable to inelastic neutron and X-ray scattering since there is a significant change in wavevector of the fast electron in addition to a change in energy.

If a sample contains dipoles, there will be a mixture of delocalized and localized phonon scattering and, as Dwyer et al. showed for BN, the dipole scattering adds a large background to the data in the typical on-axis or "bright-field (BF) EELS" (cf. Fig. 1b), which makes it very difficult to detect any localized signal from individual atoms [16]. However, the strong localization and correspondingly broad distribution in momentum space of impact scattering allows it to be separated from delocalized dipole scattering by its characteristic scattering angle.

A general way of suppressing delocalized background from dipole scattering is thus to displace the collection aperture in the so-called "dark-field (DF) EELS" geometry (cf. Fig. 1c), as first utilized by Hage et al. [17] for lattice-resolved imaging of BN. The relatively large angular displacements that were used in that study are not necessary, and limit the collected signal since the dipole scattering to be avoided is forward peaked within the range of tens of microradians beyond the forward scattering diffraction disc. As can be seen in Fig. 1c this experimental arrangement has the consequence of selecting scattering wave vectors that predominately lie in one direction which can influence the results, as discussed below.

**Demonstrations of Atomic Resolution for Vibrations**

The first demonstrations of atomic level variation in the phonon signal was by Hage et al. [17] and Venkatraman et al. [18], mapping modulations of the phonon energy-loss intensity with the spatial frequency of the crystalline lattice. Venkatraman took advantage of the fact that the two atoms in the primitive Si cell are identical so there is no dipole scattering and they were able to use an on-axis aperture (cf. Fig. 2a) while Hage et al. used DF-EELS to avoid delocalization (cf. Fig. 2b).

There have been several reports of close-to-atomic resolution vibrational EELS, obtaining vibrational information about locally different atomic structures [19-22]. However, being able to record a red-shift and suppression of bulk phonon density of states (DOS) in the vicinity of an isolated two-dimensional defect, such as the 1-layer (2.5Å) thin stacking fault in cubic SiC reported by Yan et al. in 2021, as a 7 nm wide distribution [21] is qualitatively very different from observing variations in the vibrational EELS spectrum atom column by atom column.

So far, there have been very few demonstrations of actual atomic resolution vibrational EELS. Besides the above-mentioned fundamental studies by Venkatraman et al. [18] and Hage et al. [17] in 2017, the first article to quantitatively investigate a system with significant variation of the phonon DOS on the atomic scale was published by Hage et al. in 2020 [23]. They measured EELS from a graphene sample with a point defect; a Si atom substituting one C atom (cf. Fig. 2c). An averaged spectrum from the region of the Si atom looked strikingly different from the spectrum recorded from an undisturbed graphene area, which the authors could explain in terms of localized vibrational modes of the Si atom. This study sparked more interest in this technique as it demonstrated that atomic resolution vibrational EELS was useful for imaging atomically localized modes of point defects in crystalline specimens.

Another study at atomic resolution by Xu et al. in 2023 also investigated Si point defects in graphene and compared differently coordinated Si atoms in combination with density functional theory (DFT) calculations [24]. The authors showed that it was possible to distinguish between three-fold and four-fold coordinated Si from the spectra and thus established sensitivity of the vibrational spectra in EELS to the chemical environment (cf. Fig. 2d).

The first atomic-resolution vibrational EELS study of details within extended, bulk-like defects was published by Haas et al. in 2023, where the phonon density of states (DOS) at and around grain boundaries in Si was mapped [13]. As Si has two atoms in the unit cell, it exhibits optical phonons but as there is no dipole in pure Si, there is also no delocalization, even with on-axis collection. This allows to collect the data in the bright-field geometry, and obtain a much stronger signal, as discussed above (cf. Fig. 1 b). Not only were local spectra extracted, but spectral features such as the height of the 60 meV peak were mapped at atomic resolution due to the high quality of the data (cf. Fig. 2e). The authors demonstrated significant variations in the vibrational EELS signal in 2D maps around three grain boundaries of different symmetry and obtained excellent quantitative agreement with calculations of the suppression of the phonon DOS for each one of them. They showed that special atomically-localized modes could arise, which they proposed could be used as phononic waveguides. This study demonstrates the unique capability of TEM to investigate vibrational states of bulk-like systems at atomic resolution.

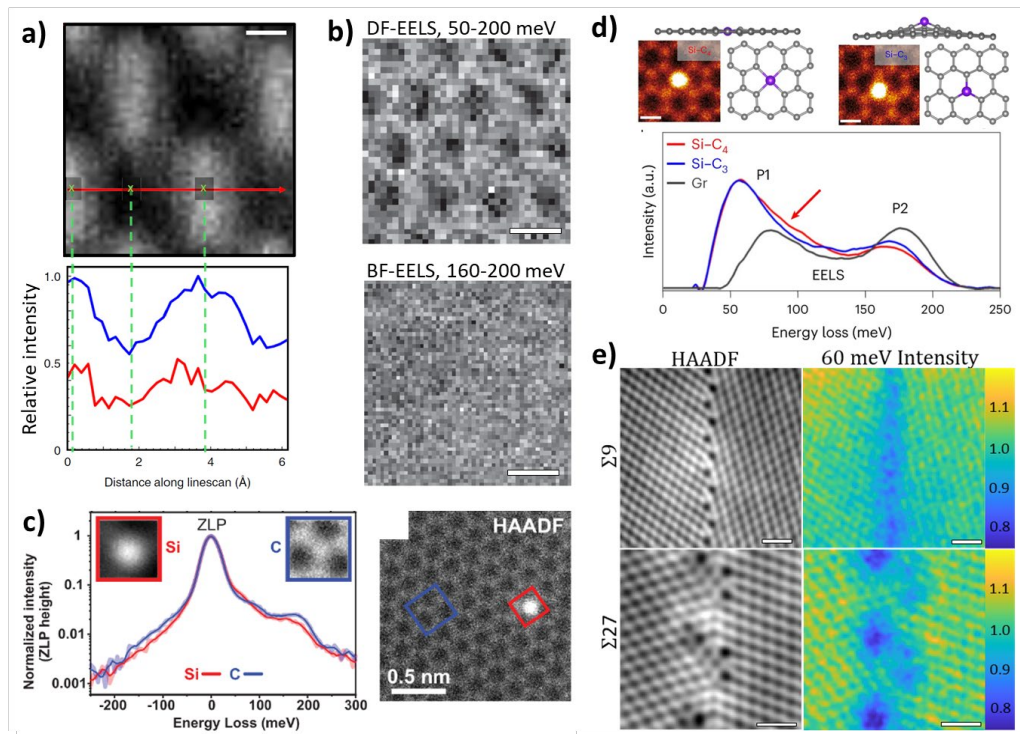

Figure 2: a) Atomic-resolution ADF (top) and profiles of the phonon intensities (bottom) showing modulation with the Si [110] lattice both for high (blue) and low (red) energies. Scale bar, 1 Å. Used with permission of Nature Research, from K. Venkatraman et al. [18]; permission conveyed through Copyright Clearance Center, Inc. b) Dark-field EELS integrating 50-200 meV showing atomic resolution for hBN (top) and no atomic contrast for 160-200 meV in BF-EELS (bottom). Scale bar,

*2 Å. Reprinted figure with permission from F. S. Hage et al. [17]. Copyright 2019 by the American Physical Society. c) Local EEL spectra from a Si substitution compared to regular graphene (left) and corresponding HAADF overview image (right). Used with permission of the American Association for the Advancement of Science, from F.S. Hage et al. [23]; permission conveyed through Copyright Clearance Center, Inc. d) Comparison of spectra of three-fold and four-fold coordinated Si defect in graphene (bottom) together with HAADF data and models of the defects (top). Scale bars, 2 Å. Used with permission of Nature Research, from M. Xu et al. [24]; permission conveyed through Copyright Clearance Center, Inc. e) HAADF images (left column) of different grain boundary types in Si [110] compared to maps of the 60 meV peak intensity (right column). Scale bars, 1 nm. Reprinted (adapted) with permission from Haas et al. [13]. Copyright 2023 American Chemical Society.*

**Theoretical Models**

This section is going into details of theoretical models to accurately account for phonon scattering in simulations and can be skipped.
Several theoretical models have been developed to describe vibrational EELS to different extent [17,25-30]. In some cases, it is sufficient to just calculate the phonon DOS [31]. The phonon dispersion across the Brillouin Zone can be constructed from the eigenvalues of the dynamical matrix at different k-points. However, this becomes impractical for systems with large numbers of atoms such as supercell models of grain boundaries. It is then more convenient to derive the phonon DOS from the Fourier transform of the time correlation between velocities of atoms of the initial time-step $t_1$ and the same atoms at subsequent times derived from molecular dynamics (MD) trajectories.

$$I(\omega_k) = \sum_{j=0}^{j=T} exp(i\omega_k t_j) \sum_{i=1}^{i=p} \sum_{n=1}^{n=N} v_{i,n}(t_j) v_{i,n}(t_1) exp\left(-\frac{\sigma^2 t_j^2}{4}\right)$$

where $v_{i,k}(t_j)$ is the difference in positions from one time step to the next and is proportional to the velocity at time $t_j$

$$v_{i,n}(t_j) = u_{i,n}(t_j) - u_{i,n}(t_{j-1})$$

$T$ is the number of time steps, $t_j$ is the $j^{th}$ time step, $N$ the number of atoms, $\omega_k$ is a frequency, $u_{i,n}$ is the coordinate of atom $n$ in the $i$'th direction (x,y,z). The total DOS is computed by summing over dimensions x, y, and z (p=3). The parameter $\sigma$ is related to the instrumental broadening, $\Delta E$ by

$$\sigma = \frac{\Delta E e}{2\hbar\sqrt{ln2}}$$

when $\Delta E$ is in eV. To derive meaningful frequencies from MD simulations a good empirical potential is needed. Tersoff gives a very reliable empirical potential for Si and SiC that was used in calculations of phonon DOS localized at the cores of grain boundaries in Si [32] and in the comparison with experimental measurements [13]. For other materials, especially those with a high degree of ionicity, good empirical potentials are much harder to find. However, in a recent paper machine learning was used to derive a potential for $SrTiO_3$ with a training set derived from DFT calculations of forces on atoms with a range of displacements [33]. In some ways this is comparable to what Radtke did in piecing together a dynamical matrix for a large supercell from DFT calculations for a perfect crystal and small regions around a defect [23].

To quantitatively compare measured and calculated intensities it is also necessary to take into account dynamical diffraction, i.e. multiple elastic scattering of the fast electron, in addition to the phonon scattering. We will focus on two frameworks here that allow for energy-resolved data and include dynamic diffraction and are thus very relevant to model versatile atomic-resolution vibrational EELS experiments. In crystals the elastic scattering is concentrated in the directions allowed by Bragg scattering, this concentration means that even for modest crystal thicknesses it is necessary to take account of multiple elastic scattering, which is also known as dynamical diffraction. In comparison, phonon scattering is weak and can be treated as a first order perturbation.

The first model is an extension of the frozen phonon based multislice using MD to provide configurations of atomic coordinates in a supercell and is called "frequency-resolved frozen phonon multislice" (FRFPMS), which was developed by Zeiger and Rusz [34-36]. The second model was developed by Rez and Singh [37] and uses phonon dispersions from DFT or MD calculations in a Bloch wave-based approach. We will briefly outline the first model before describing the second framework in some more theoretical detail. An exhaustive list of other existing models and discussion of some of them was recently provided by Zeiger et al. [30].

FRFPMS contains no dipole scattering and therefore only yields the impact scattering part. Thus, it is only suitable for dipole-free systems or for DF-EELS where the dipole scattering is avoided due to the detection geometry. As these are the cases relevant for atomic-resolution vibrational EELS, this approach is generally very suitable in the context of our discussion. Various "snapshots" are taken of all the atomic positions e.g. from a frequency-resolved MD run, typically over the course of 250 ps with a 0.5 fs time step [35]. These N snapshots are used as inputs for a full multislice calculation of the exit surface wavefuction $\varphi$ which is a function of probe position $r_p$ and atom coordinates $R_n$. The vibrational or phonon scattering is the difference between the incoherent scattering $I_{inc}(\mathbf{q},\mathbf{r}_b,\omega)$ and the coherent elastic scattering $I_{coh}(\mathbf{q},\mathbf{r}_b,\omega)$.

$$I_{inc}(\mathbf{q},\mathbf{r}_b,\omega) = \frac{1}{N}\sum_{n=1}^{N}|\varphi(\mathbf{q},\mathbf{r}_b,\mathbf{R}_n(\omega))|^2$$

$$I_{coh}(\mathbf{q},\mathbf{r}_b,\omega) = \left|\frac{1}{N}\sum_{n=1}^{N}\varphi(\mathbf{q},\mathbf{r}_b,\mathbf{R}_n(\omega))\right|^2$$

$$I_{phonon}(\mathbf{q},\mathbf{r}_b,\omega) = I_{inc}(\mathbf{q},\mathbf{r}_b,\omega) - I_{coh}(\mathbf{q},\mathbf{r}_b,\omega)$$

This is very similar to the procedure used by Hall and Hirsch [38] to calculate TDS. The advantages of this method is that it can be used on large supercells such as an antiphase boundary in hBN [36] and that it also includes anharmonic terms. The disadvantage is that it is very compute intensive for perfect crystals, especially if one wants to calculate an image that involves running full multislice calculations for a large supercell for each probe position. However, computation time is independent of – and thus not increasing with a reduction of – symmetry. Originally the frequency selection was performed by means of a thermostat in the MD that only allowed for phonon modes in a restricted energy range. The method has since been revised to a more efficient approach by bandpass-filtering the Fourier transforms of MD trajectories [39].

The second theoretical framework that we will discuss here is based on expressing the phonon scattering as a sequence of operators for dynamical elastic diffraction of the incident electrons, followed by phonon scattering and further dynamical diffraction of the phonon scattered electrons [37]. As this model includes impact as well as dipole scattering, it is able to calculate not only the DF- but also the BF-EELS signal for materials containing dipoles, in contrast to FRFPMS. However, in its Bloch wave-based implementation, the model is not able to efficiently describe non-periodic structures as this would require a prohibitive number of Bloch waves.

In this framework, wavevectors are broken down into a component in the 1st Brillouin Zone, labeled **q** for incident electrons and **q'** for phonon scattered electrons, and a Bragg vector labeled by lower case letters, **g,h** for the incident electron and uppercase letters **L,M** for the phonon scattered electrons, see Fig. 1e. The phonon scattered amplitude from a slice at depth z arising from incident electrons in a direction specified by q+h is

$$\psi(\mathbf{h}, \mathbf{q}' - \mathbf{q}, t, z) = \sum_{M,L,g} R_{ML}(\mathbf{q}', t-z) H(\mathbf{q}' + \mathbf{L} - \mathbf{q} - \mathbf{g}) P_{gh}(\mathbf{q}, z) A(\mathbf{q} + \mathbf{h})$$

where $A(\mathbf{q+h})$ defines the incident probe, $P_{gh}(\mathbf{q},z)$ represents dynamical diffraction of the incident electrons to a depth z where they are phonon scattered represented by the operator $H(\mathbf{q'+L-q-g})$ and then these electrons can undergo further dynamical diffraction represented by $R_{ML}(\mathbf{q'},t-z)$. The phonon scattering operator, which depends on a wavevector $\mathbf{Q=q-q'}$ inside the first Brillouin Zone and a Bragg vector $\mathbf{G=L-g}$, can be derived by taking the displacement from the phonon as a small perturbation in the expression for the electron scattering factor

$$H(\mathbf{Q} + \mathbf{G}) = \left(1 + N\left(\frac{\hbar\omega(Q)}{k_B T}\right)\right)^{\frac{1}{2}} \sum_j (\mathbf{Q} + \mathbf{G}) \cdot \mathbf{e}_j \left(\frac{\hbar}{2m_j\omega(Q)}\right)^{\frac{1}{2}} f_{el}^j(Q + G) exp(i\mathbf{Q} \cdot \mathbf{R}_j)$$

where **e** is the polarization vector $f_{el}^j$ is the electron scattering factor for atom *j* and $\mathbf{e}_j$ is the direction of atom movement extracted from the eigenvector of the dynamical matrix. The sum of the product of electron scattering factors and atomic displacements resolved along the scattering wave vector direction can lead to systematic absences [40].

If there is no scattering from one Brillouin Zone to the next and the Bragg vector is zero, the phonon scattering is called normal. When the phonon scattering crosses a Brillouin Zone boundary it is called Umklapp scattering. As the strength of phonon scattering depends on the product of scattering wavevector and electron scattering factor, Umklapp scattering from the 1st to the 2nd or even 3rd Brillouin Zone dominates – this should be considered when trying to optimize the signal collection geometry. This effect can be seen in experimental data depicted in Fig. 1d; while the diffraction intensity (bottom) quickly diminishes with scattering angle, the phonon scattering first becomes stronger before then diminishing slowly.

Lattice resolution comes about from constructive interference between directions in the incident beam separated by a Bragg angle.

$$I(\mathbf{r_p}) = \sum_{h,h'} \sum_{Q,z} \psi(\mathbf{h}, \mathbf{Q}, t, z) \psi^*(\mathbf{h'}, \mathbf{Q}, t, z) \, exp(i(\mathbf{h} - \mathbf{h'}) \cdot \mathbf{r_p}) \Delta z$$

To calculate the intensities in the phonon scattered image it is necessary to calculate the dynamical diffraction propagation matrices $P_{gh}(\mathbf{q},z)$ and $R_{ML}(\mathbf{q'},t-z)$. This could be done using Bloch waves where the dependence on depth is already explicit in the complex exponential, or by multislice methods with a supercell where the complete wavefunction for each depth would be stored. This would then have to be partitioned into a part that relates to **q** or **q'**, the wavevector component in the 1st Brillouin Zone, and the Bragg vector. A very large supercell would be needed for a fine grid in phonon wavevector. The problem with Bloch wave methods is that since diagonalisation scales as $N^3$ they are restricted to perfect crystal systems that can be represented by a small number of beams or Fourier coefficients of potential. Whichever technique is used there is elaborate "book-keeping" to track the Umklapp terms and make sure that a scattering wavevector reduced to the 1st Brillouin Zone is available for calculation of frequency for a relevant branch of the phonon dispersion surface. For systems with a small number of atoms, this is conveniently done by diagonalising the dynamical matrix calculated from standard DFT codes such as VASP or Quantum Expresso as implemented in Phonopy [41].

**Other Techniques**

While inelastic neutron and more recently inelastic X-ray [42] scattering techniques are able to measure the dispersion of phonons, they lack the spatial resolution for atomic-scale vibrational analyses. Only tip-enhanced optical spectroscopy, namely tip-enhanced Raman spectroscopy (TERS), has been able to achieve spatial resolution of vibrational modes on the Ångström-scale in a few instances [43]. While TERS is mostly restricted to surfaces, the electron microscope is still, also in the case of vibrational states, the only instrument capable of atomic resolution for bulk-like specimens. Another difference is that, while infrared- and Raman- active modes are mutually exclusive in light optics as they depend on polarization and polarizability, respectively, the electron can excite both types as discussed above. Moreover, electrons can transfer much larger momenta for a given energy than photons and thus probe vibrational states that are "dark", to optical spectroscopy, i.e. not directly measurable by optical means. An advantage of TERS is that it does not rely on fast electrons and can therefore better probe beam-sensitive materials at high resolution. As EELS probes the projected DOS through the whole sample and TERS has a high degree of surface sensitivity, we propose that they could be used in combination to acquire complementary data allowing to separate surface and bulk effects. Moreover, tip-enhanced techniques can probe vibrations at resolutions limited by the tip apex but only close to zero momentum, while EELS can be used in DF-geometry to avoid delocalization but then probes only higher momenta. This also indicates a complementarity of the techniques that could be exploited.

**Future Perspectives**

To optimize the acquisition of the phonon signal by EELS, a large collection angle of the spectrometer is useful, which means that correction of spectrometer aberrations to higher orders is needed. To exclude the BF signal, the direct beam could be blocked, by means of a beam stop available in most microscopes in combination with a suitable camera length (magnification of the diffraction pattern). So far this has not been done. Another option would be to have an annular spectrometer entrance aperture. However, these approaches have the disadvantage that they necessitate the collection of a wide angular range by the spectrometer, which places strong constraints on spectrometer aberration correction to still achieve the necessary energy resolution. The advantage of DF-EELS is that the necessary angular range that needs to be collected can be relatively small if a tilt is introduced between the scattering from the sample and the spectrometer entrance aperture. However, this only collects a fraction of the azimuthal angular range for the given radial angular range.

As the phonon signal is low due to the small fraction of current passed by the monochromator, it is typical to use larger beam currents even for optimized collection geometries to generate sufficient signal to image the atomic structure on the time scale of a few seconds needed to efficiently navigate to the region of interest and optimize acquisition conditions while keeping it centered in the presence of typical drift rates. However, since this increased initial current is obtained by a lower demagnification of the source, this means that the effective source size is increased and the spatial coherence reduced. Reducing the beam current by monochromation has no (beneficial) influence on the source size. In atomic-resolution vibrational EELS experiments, this typically becomes the main limitation to spatial resolution rather than aberrations that gain importance at the high convergence angles needed for a small probe size and spectrometer aberrations that can now be corrected. In comparison, in non-monochromated STEM experiments, the aberration-correction of the convergent probe is the main limitation to spatial resolution.

The brightness-limited signal makes it in practice necessary to integrate the EELS intensity for an extended time to obtain a sufficient signal-to-noise ratio. This leads to noticeable drifts for atomic-resolution EELS, even for very stable sample stages. A way to remedy this is the registration of STEM data by acquiring multiple data sets from the same area and correcting for drift and scan artifacts using the supposed redundancy of the data and some prior knowledge of the serial acquisition process. Non-rigid registration of spatially-resolved STEM-EELS has been demonstrated [44] and was later extended to series of four-dimensional STEM data sets [45], which is relevant for the following discussions.

The detector is critical for acquiring high quality data. With the advent of hybrid-pixel direct electron detectors it has become possible to acquire data with no readout-noise at high frame rates and relatively high currents. This makes it possible to acquire the zero-loss peak in combination with other, much weaker, parts of the spectrum [46] and additionally use very short dwell times suitable for atomic resolution acquisition and then to utilize the registration scheme mentioned above without accumulating readout-noise. This combination of detector and acquisition/processing scheme has led to high fidelity atomic resolution vibrational EELS data [13].

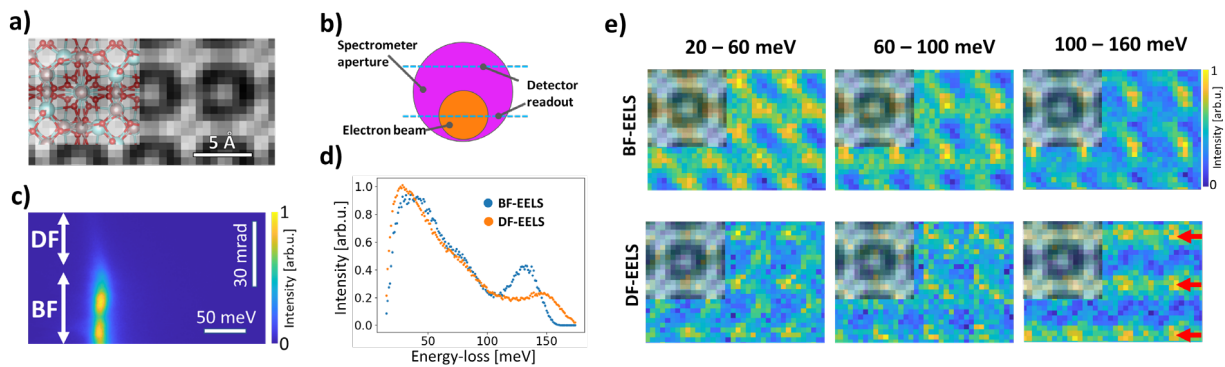

Figure 3: a) HAADF image of GAGG crystal in [001] zone axis overlayed with structure model. b) Sketch of the relative sizes and positions of electron beam, spectrometer aperture and camera readout. The electron beam width (full convergence angle) is 60 mrad. c) Single detector frame corresponding to one real space pixel with the strong intensity stemming from the elastically scattered electrons at zero energy-loss. d) Averaged background-subtracted spectra for the dark-field and bright-field regions. e) Extracted spatial maps for different energy windows and for BF and DF regions as indicated. Especially in the DF map for 100-160 meV it can be seen that the 4-fold symmetry is reduced to a 2-fold symmetry and the atoms along the rows indicated by the arrows stand out with stronger contrast.

Although Heisenberg's uncertainty principle does not allow for sufficiently fine momentum information at atomic resolution to investigate the dispersion of phonon, additional momentum information is still very valuable for atomic resolution vibrational EELS. Fig. 3 demonstrates parallel BF- and DF-EELS mapping at atomic resolution for a gadolinium aluminum gallium garnet (GAGG) crystal in [001] zone axis. Fig. 3a depicts an HAADF image acquired in parallel with the spectroscopy data, overlaid with a structure model. To enhance the signal, the data was averaged over identical unit cells by means of template-matching. A sketch of the collection geometry is shown in Fig. 3b with the sizes of the EELS aperture, the convergence angle and the range of momentum transfer recorded by the detector. The experimental parameters were 30 mrad semi-convergence angle, 60 keV primary electron energy, 80 pA initial beam current monochromated to 4 pA, 62.5 pm/px real space sampling. The detector spans bright and dark field regions of similar size. Fig. 3c shows a single detector frame with the zero-loss peak clearly visible in the bright-field region and much weaker in the dark-field. In Fig. 3d averaged and background-subtracted spectra for the DF and BF regions are shown. A comparison of BF and DF for different energy windows is shown in Fig. 3e. It should be noted that the reduced initial current (before monochromation) reduces the source size and leads to a spatial resolution of about 1.5 Å. This presents a pathway to achieving even finer details in vibrational EELS.

In Fig. 3e it can be seen that the atoms primarily associated with certain vibrational energies have stronger intensities in the generated real space images. As the dipolar part is not dominating the BF signal due to the low polarity of bonds in GAGG (and there is also some DF signal mixed in, cf. Fig. 3b), the BF still shows variations within the unit cell. An interesting feature of the data is that it shows breaking of the four-fold symmetry in real space compared to the HAADF data (and structural model) resulting in a 2-fold symmetry (highlighted by the arrows). This asymmetry is due to the off-axis tilt in the collection geometry selecting a direction for the scattering wave vector and enhancing the contribution for vibrational modes in that direction. Recently, Yan et al. demonstrated in an experiment with two displacement directions of the DF-EELS that directional momentum information can be extracted from such data in agreement with calculations [39].

Atomic resolution vibrational EELS with additional momentum axes would lead to very rich data sets that should prove useful for detailed investigations of vibrational phenomena at the atomic scale. A scheme of how to obtain energy- and momentum-resolved data at the atomic scale was recently demonstrated by some of the authors for core-loss excitations [47] and can readily be applied to vibrational EELS. Full detector frames of energy-loss and momentum-transfer in one direction are recorded for each probe position, just like in Fig. 3. Acquiring multiple maps for the same probe position but with differently rotated diffraction patterns in the spectrometer entrance plane allows five-dimensional datasets $I(x,y,k_x,k_y,E)$ to be recorded, i.e. sampling the momentum plane instead of just one momentum axis. The aforementioned registration scheme for multi-dimensional data [45] facilitates the spatial coincidence with atomic precision of this series of four-dimensional data sets and thus allows to integrate the series dimension, therefore obtaining two momentum dimensions in addition to the spatial and energy dimensions. This was shown for the simplest case of two perpendicular momentum projections [47], but it would also be possible to sample the rotation more finely, leading to a finer reconstruction of momentum plane. This scheme would also allow to collect more dark-field signal without the need for a special annular aperture.

Another future perspective is the combination of TERS with vibrational spectroscopy which could be very powerful, as mentioned above. Combining STEM-EELS with a surface-sensitive technique would allow to separate signals into bulk and surface contributions. In the case of very beam-sensitive materials, an investigation by TERS prior to vibrational EELS would offer some non-destructive high-resolution vibrational information, e.g. about molecules on a surface, before then using the damaging but information-rich electron scattering to study momentum-dependence or Raman-inactive modes.

Recently, it has been shown that the creation of phonon-polaritons, which arise in some materials, can play a significant role in DF-EELS due to multiple scattering [48]. This finding again underlines the importance of detailed simulations to fully understand experimental results. As the quality of experimental data is significantly increasing, it will be relevant to have theoretical models capable of describing atomic-resolution phonon EELS that explicitly include higher order terms like double phonon scattering. In this context, a theoretical method based on time-autocorrelations of auxiliary wavefunctions was recently proposed, which allows to include multi-phonon excitations to any order in addition to dynamical diffraction and anharmonicities [49].

As there seems no obvious physical limitation, future monochromator and spectrometer designs might push the energy resolution to well below 1 meV. This will allow for the investigation of materials made of heavier atoms, or featuring weaker bonds, and better separation of closely spaced spectral features.

ACKNOWLEDGEMENTS

The authors thank Paul Zeiger and José Ángel Castellanos-Reyes (Uppsala University) for comments on the manuscript and Niklas Dellby (Bruker AXS) for advice. Alisa Ukhanova (Lomonosov University) and Oleg Busanov (Fomos-Materials) is thanked for providing the GAGG sample. B.H. acknowledges the German science foundation (DFG) for financial support in the framework of project 530143441 "PuMMAVi".